\documentclass[twocolumn]{aastex62}
\usepackage{natbib}
\newcommand{\rave}{\textsc{Rave}\ }

\received{\today}
\accepted{TBD}

\submitjournal{AJ}

\shorttitle{Single-lined spectroscopic binaries from \textsc{Rave} and \textsc{Gaia} DR2}
\shortauthors{Danijela Birko, et al.}

\begin{document}

\title{Single-lined Spectroscopic Binary Star Candidates from a Combination of the \rave and \textsc{Gaia} DR2 surveys}

\correspondingauthor{Danijela Birko}
\email{danijela.birko@gmail.com}

\author[0000-0002-5571-5981]{Danijela Birko}
\affil{University of Ljubljana, Faculty of Mathematics and Physics, Ljubljana, Slovenia}

\author[0000-0002-2325-8763]{Toma\v{z} Zwitter}
\affil{University of Ljubljana, Faculty of Mathematics and Physics, Ljubljana, Slovenia}

\author[0000-0002-1891-3794]{Eva K.\ Grebel}
\affil{Astronomisches Rechen-Institut, Zentrum f\"ur
Astronomie der Universit\"at Heidelberg, M\"onchhofstr.\ 12--14,
69120 Heidelberg, Germany}

\author[0000-0001-6516-7459]{Quentin A Parker}
\affil{CYM Physics Building, The University of Hong Kong,  Pokfulam, Hong Kong, SAR, PRC}
\affil{The Laboratory for Space Research, Hong Kong University, Cyberport 4, Hong, Kong, SAR, PRC}

\author[0000-0002-9035-3920]{Georges Kordopatis}
\affil{Universit\'e C\^ote d'Azur, Observatoire de la C\^ote d'Azur, CNRS, Laboratoire Lagrange, France}

\author[0000-0001-7516-4016]{Joss Bland-Hawthorn}
\affil{Institute of Astronomy, School of Physics, University of Sydney, Australia}

\author[0000-0001-6280-1207]{Kenneth Freeman}
\affil{RSAA Australian National University, Camberra, Australia}

\author[0000-0002-1317-2798]{Guillaume Guiglion}
\affil{Leibniz-Institut f\"{u}r Astrophysik Potsdam (AIP), An den Sternwarte 16, D-14482 Potsdam, Germany}

\author[0000-0003-4446-3130]{Brad K. Gibson}
\affil{Jeremiah Horrocks Institute for Astrophysics \&\  Super-computing, University of Central Lancashire, Preston, UK}

\author{Julio Navarro}
\affil{CIfAR Fellow, University of Victoria Physics and Astronomy, Victoria, BC V8P 5C2, Canada}

\author{Warren Reid}
\affil{Department of Physics and Astronomy, Macquarie University, Sydney, NSW 2109, Australia}
\affil{Western Sydney University, Locked bag 1797, Penrith South, NSW 2751, Australia}

\author[0000-0003-4072-9536]{G. M. Seabroke}
\affil{Mullard Space Science Laboratory, University College London, Holmbury St Mary, Dorking, RH5 6NT, UK}

\author[0000-0001-6516-7459]{Matthias Steinmetz}
\affil{Leibniz-Institut f\"{u}r Astrophysik Potsdam (AIP), An den Sternwarte 16, D-14482 Potsdam, Germany}

\author[0000-0002-3590-3547]{Fred Watson}
\affil{Anglo-Australian Observatory, Sydney, Australia}

\begin{abstract}

The combination of the final version of the
\rave spectroscopic survey data release 6 with radial velocities and astrometry from \textsc{Gaia} DR2 allows us to identify and create a catalog of single-lined binary star candidates (SB1), their inferred orbital parameters, and to inspect possible double-lined binary stars (SB2).

A  probability function  for the detection  of  radial  velocity  (RV)  variations is  used  for  identifying  SB1 candidates. The estimation of orbital parameters for main sequence dwarfs is performed by matching the measured RVs with theoretical velocity curves sampling the orbital parameter space.  The method is verified  by  studying  a  mock  sample  from  the  SB~9 catalog. Studying the boxiness and asymmetry of the spectral lines allows us to identify possible SB2 candidates, while matching their spectra to a synthetic library indicates probable properties of their components.

From the RAVE catalog we select 37,664  stars  with  multiple  RV  measurements  and  identify  3838 stars as SB1 candidates.  Joining \rave and \textsc{Gaia} DR2 yields 450,646 stars with RVs measured by both surveys and 27,716 of them turn out to be SB1 candidates, which is an increase by an order of magnitude over previous studies. For main sequence dwarf candidates we calculate their most probable orbital parameters: orbital periods are not longer than a few years and primary components have masses similar to the solar mass. All our results are available in the electronic version. 
\end{abstract}

\keywords{binaries: spectroscopic --- surveys --- methods: data analysis}

\section{Introduction} \label{Sec:introduction}
The majority of stars are members of multiple systems of two or more gravitationally bound stars. In the vast majority of cases these stars are coeval and have an identical chemical composition, and in favorable cases their masses and/or sizes can be determined directly. Yet, in the majority of systems with orbital periods of weeks to years it is a challenge even to identify their multiple nature. Detection of variability of radial velocities (RVs) is a very successful, though a relatively time-consuming method, which has been extensively used in studies of dwarfs and sub-dwarfs \citep{duquennoy1991,fischer1992}, massive binary systems 
\citep[e.g.][]{sana2009} and binary stars in clusters \citep{abt1999,sommariva2009}. There have been several surveys dedicated to the search of spectroscopic binaries \citep[e.g.][]{latham2002,griffin2006,mermilliod2007}. The Geneva-Copenhagen Survey \citep{nordstrom2004} marked a milestone in the size of the sample, as it presented $\sim 5$ RV measurements of 14,139  F- and G-type dwarfs drawn from a kinematically unbiased magnitude-limited sample, painstakingly observing one star at a time. A landmark result of this effort was the realization that spectroscopic binarity can be detected in 19\%\ of the observed targets, while the fraction of binary stars of all types reached 34\%, in agreement with an earlier result of \citet{duquennoy1991}.   

The last decade has been marked by much larger ground-based  spectroscopic surveys, which use wide-field coverage and fiber optics to obtain spectra of a hundred or more stars at a time. The results are not only RVs but also spectroscopically determined values of stellar parameters, including chemistry. On the other hand, the scientific focus is shifting from stellar kinematics to Galactic archaeology, so the goal is to observe as many stars as possible. This means that most targets are observed only during a single night, with the majority of repeated observations consisting of two visits per target scheduled days to years apart. So any statement on binarity from these spectroscopic surveys is based on a large number of targets with a small number of visits. These properties are typical for the \rave (Radial Velocity Experiment) survey \citep{steinmetz2006}, but also for \textsc{Gaia-ESO} \citep{gilmore2012}, \textsc{Apogee} (Apache Point Observatory Galactic Evolution Experiment) \citep{holtzman2015}, \textsc{Lamost} (The Large Sky Area Multi-Object Fibre Spectroscopic Telescope) \citep{liu2017}, and \textsc{Galah} (The GALactic Archaeology with HERMES) \citep{desilva2015} surveys. 

Here we focus on spectroscopic binaries that can be identified in the \rave survey. Double-lined binaries (SB2) where we identify both sets of spectral lines are quite rare \citep{matijevic2012}, as they imply a very similar mass of both components. Another major reason why SB2 are rare is that the geometry needs to be rather favorable for us to see the line split. So our primary goal is to identify single-lined binaries (SB1), binaries where only spectral lines of primary components can be detected, which are much more common. We base our approach on an earlier study \citep{matijevic2011}, but with two important upgrades: (i) our analysis is based on the final and complete set of \rave spectra \citep{steinmetz2019} which approximately doubles the considered sample, (ii) RVs derived by \rave are matched to those of ESA's \textsc{Gaia} space mission \citep{gaia2018} which increases the sample with multiple measurements by an order of magnitude. A Monte-Carlo approach is used to infer physical properties of the identified SB1 binaries. Their spectra are also searched for the presence of light from a secondary component. 

We start with a sample based on \rave observations only. In Sec.~2 we present the data and in Sec.~3 we summarize the method to select SB1 candidates. Next, we present basic properties of the SB1 sample, statistical inference of values of fundamental parameters, and results of the search for absorption lines from a secondary component.
In Section 7  we repeat the whole process, now including also RV observations from the \textsc{Gaia} satellite which are, however, different enough from \rave observations to keep their analysis separate from the one using \rave observations only. Finally we add some discussion of the results and an outline for the future. 

\section{\rave observations and sample selection} \label{Sec:raveobs}

The RAdial Velocity Experiment \citep{steinmetz2006,zwitter2008,siebert2011,kordopatis2013,kunder2017,steinmetz2019} is a medium resolution (R $ \sim $ 7500) spectroscopic survey of the Milky Way. It used the UK Schmidt telescope at the Australian Astronomical Observatory to obtain over half a million stellar spectra over the period of 12 April 2003 to 4 April 2013. These cover wavelength range of 8410 -- 8795 \r{A}. The survey properties as well as all its data products and analysis are described in detail in its final data release paper \citep{steinmetz2019}. Here we provide just a brief summary as a service to the reader. 

\rave is the first systematic (wide field coverage) spectroscopic Galactic archaeology survey. While the survey was ongoing its goals were gradually surpassing its original name by supplementing determination of radial velocity with estimates of effective temperature,  surface gravity, and chemical properties, including abundances of aluminum, iron, magnesium, silicon, titanium and nickel in the stellar photospheres (these abundances are quoted in order of increasing uncertainties, which generally range from 0.14 to 0.23 dex). An inclusive approach has been used, where information obtained from observed spectra was supplemented with complementary photometric and astrometric information, as it became available. 

The final \rave data release contains 518,392 spectra of 451,788 stars, which present a magnitude limited sample with 9~$<$ \textit{I} $<$~12. The typical signal-to-noise ratio (S/N) of the measured spectra is $ \sim $~40 per pixel. The \rave wavelength range matches that of the \textsc{Gaia} mission. This wavelength range includes a lot of spectral lines; most importantly the singly ionized calcium triplet ($ \lambda \lambda $ = 8498, 8542, 8662 \r{A}), the Paschen series of hydrogen, and Fe I multiplets. In the measured part of the spectrum, contributions from telluric lines can be neglected and the only significant spectral signature of the interstellar medium is the diffuse interstellar band at 8620 \r{A} \citep{munari2008,kos2014}. 

The selection of the \rave targets was very close to a random magnitude-limited sample of southern stars, but avoiding fields closer than $\sim 5$ degrees from the Galactic plane and those in the direction of the Galactic bulge. Details of the selection function are discussed in \citet{wojno2017}. A random selection implies that some of the stars belong to rare spectral types or brief evolutionary stages. Local linear embedding was shown to be an efficient morphological classification technique to pinpoint such peculiar cases and has been applied to \rave \citep{matijevic2012}. While morphological classification proved efficient in detecting SB2 objects and chromospherically active stars it is clear that it cannot identify SB1 stars which are hidden among the vast majority of 90-95\%\ of stars with morphologically normal spectra.

\begin{figure}
    \centering
    \includegraphics[width=\columnwidth]{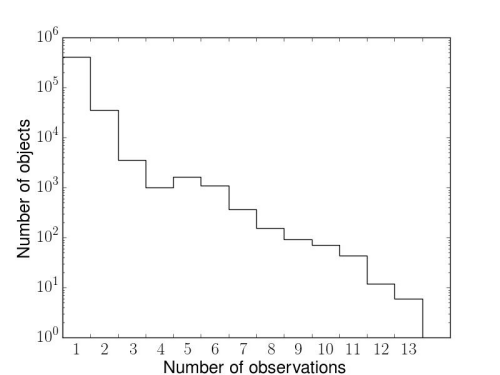}
    \caption{Histogram of the number of RAVE observations per object.}
    \label{Fig:noobs}
\end{figure}

\begin{figure}
   \centering
   \includegraphics[width=\columnwidth]{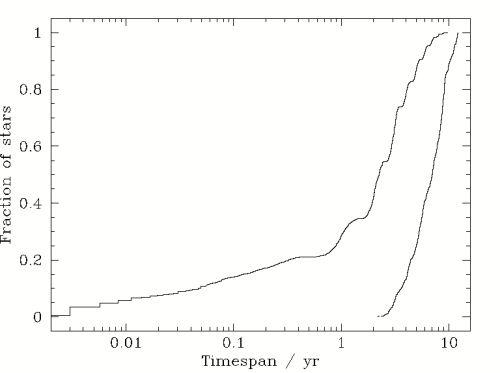}
   \caption{Cumulative plot of a time-span between the first and the last observations of the same object within the \rave survey (left line) and for a combination of \rave data with RV measurements of \textsc{Gaia}  (right  line). In the latter case the assumed epoch for the \textsc{Gaia} observation is 15 June 2015.}
   \label{Fig:timespan}
\end{figure}

The derivation of RVs is the main result of interest to us here. Velocities are derived as described in \citet{siebert2011}. A two stage process is used. First a rough estimate of RV,  with a typical precision better than 5 km s$^{-1} $, is obtained using a subset of 10 template synthetic spectra covering a wide range of stellar parameters. Next, a best-matching template is constructed using the full template database with a penalized chi-square technique described in \citet{zwitter2008}. This template then allows us to determine the final, more precise RV, which is corrected for possible zero-point shifts (due to thermal instabilities of the instrument) and reported in the inertial frame of the Solar barycenter. As discussed in \citep{steinmetz2019}, the typical error of the derived RV in the \rave survey is $\sim 1.1$~km~s$^{-1}$.

SB1s can be identified from multiple RV measurements of good quality, so we used the following selection criteria:
\begin{enumerate}
\item to safeguard against systematic problems with measurements of noisy spectra we used only those with a SNR\_avg\_SPARV $\geq 20$  in \rave DR6;
\item only stars that have all their spectra classified as normal (flag1 with the value 'n') were used in \rave DR6;
\item only stars with at least two RV determinations were considered. 
\end{enumerate}
The first criterion is fulfilled for 414,637 (91.8~\%) stars and the second further narrows selection to 395,919 (87.4~\%) stars in RAVE DR6. The third criterion is more selective, though unavoidable in a search of SB1 candidates.  In total 47,360 stars (9.1~\%) have multiple observations. When applying the other two criteria we end up with a sample of 37,661 stars with repeated observations. 
Some of the targets have been observed up to 13 times (Figure \ref{Fig:noobs}). Most of the stars with at least 6 visits are located at Galactic latitude $b > 30^o$ and are part of a logarithmic cadence with observations separated by approximately 1, 4, 10, 40, 100, and 1000 days. But a vast majority of stars have only two spectra, form a random sub-sample of the RAVE survey and are of primary interest to us here. Some of the repeats were made only days apart but about half have a time-span longer than 2 years (Figure \ref{Fig:timespan}). 

\section{The Method} \label{Sec:method}

\begin{deluxetable*}{rrrrrrrr}
\caption{Number and fraction of SB1 candidates for different values of $ p_{log} $. $N$ is the number of objects with $Nobs$ observations per object. Fraction of SB1 is higher for higher number of observations. Longer time span between re-observations of objects with higher $Nobs$ results  in a higher percentage of SB1s. 
\label{Tab:noobs}}      
\tablehead{
$Nobs$ & $N$ & \multicolumn{2}{c}{$p_{log} > 2.87$} & \multicolumn{2}{c}{$p_{log}> 4$} & \multicolumn{2}{c}{$p_{log} > 6$}\\
  	& 	&	$N$ 	& \% 	& $N$ 	& \% 	& $N$ 	& \%\\
}
\startdata
2	&	31059	& 2384	& 7.7	& 1694	& 5.5	& 1183	& 3.8	\\
3	& 	2744	& 394	& 14.4	& 273	& 10.0	& 166	& 6.0	\\
4	& 	943		& 182	& 19.3	& 107	& 11.3	& 61	& 6.5	\\
5	& 	1269	& 276	& 21.7	& 166	& 13.1	& 102	& 8.0	\\
6	& 	1015	& 326	& 32.1	& 210	& 20.7	& 106	& 10.4	\\
7	& 	345		& 120	& 34.8	& 76	& 22.0	& 43	& 12.5	\\
8	& 	131		& 60	& 45.8 	& 39	& 30.0 	& 24	& 18.3	\\
9	& 	53		& 23	& 43.4	& 16	& 30.2	& 8 	& 15.1	\\
10	& 	60		& 33	& 55.0	& 28	& 46.7	& 16	& 26.7	\\
11	& 	35		& 19	& 54.3	& 16	& 45.7	& 8		& 22.9	\\
12	& 	7		& 3		& 42.9	& 2		& 28.6	& 0		& 0.0	\\
$\geq$ 2	& 	37661 & 3838 & 10.2	& 2627	& 7.0	& 1717	& 4.6	\\
\enddata
\end{deluxetable*}

\quad The identification of SB1 candidates is based on the detection of their RV variability. So we need a quantitative criterion for considering changes in RV as significant. Following the method and reasoning from \citet{matijevic2011}, one can write the probability that $RV_2$ is larger than $RV_1$ as 
\begin{equation}
{P(2>1)} = \frac{1}{2} \left[ 1 + \text{erf}\left(\frac{\text{RV}_{2} - \text{RV}_{1}}{\sqrt{2(\sigma_{1}^{2} + \sigma_{2}^{2})}} \right) \right]
\end{equation}

\noindent where RV$ _{1} $ and RV$ _{2}$ are radial velocities and $ \sigma_{1} $ and $ \sigma_{2} $ errors measured for the same star at the two different observations. The squares of the RV errors $ \sigma_{i}^{2} $ can be treated as variances of the Gaussian distribution with the RV$_{i}$ as the mean value. If we would pick two samples from each of these distributions, $P$(2$>$1) represents the probability that the pick from the second sample is greater than the pick from the first one. If the RVs are the same, the numerator of the error function will be zero and the probability will equal 1/2. For a pair of very different RVs and comparably small errors, the error function approaches 1, and consequently the complete probability goes to 1. For stars with a significant RV variability the value of $P$ should be close to 1, so we introduce a new function that includes a logarithm of $P$
\begin{equation}
p_{\text{log}} = -\text{log}_{10}(1-P)
\end{equation}

\noindent For objects with very significant $RV$ variability the argument of the logarithm can be very small and cause floating point errors, so we limit the value of $ p_{log} $ to 14.
\citet{pourbaix2005} use $ p_{log} $ = 2.87 as a lower limit indicating a significant RV variability, assuming equal RV errors. Such a limit on $p_{log}$ corresponds to RV values that are $4.24\, \sigma_{i}$ apart. Similarly, $p_{log}  < 2$ corresponds to RV values less than $3.3\, \sigma_i$ apart, so the variability is questionable. And $ p_{log} < 1$ implies RV differences smaller than $1.8\, \sigma_i$, so an insignificant RV variability.

Detection of RV variability is not a sufficient criterion to identify a SB1 object. We want to check if RV variability is not caused by surface activity and if the object shows photometric variability that is unlikely considering long orbital periods of a majority of SB1 stars. Although we required that spectra of SB1 candidates are morphologically classified as those of normal single stars we made additional cross-checks. In particular, \citet{zerjal2013} made a catalog of chromospherically active stars, and photometric variability can be identified using the Rave DR5 $+$ Gaia DR2 photometic variability flag (phot\_variable\_flag). These checks do not change our results significantly. Among the 3838 candidates discussed in the next section only 3 are known to be chromospherically active and 17 have a flag for photometric variability, most of them are red giants. Also, we checked RV variability as a function of time span. Close or semi-detached binaries that fill their Roche lobe, with periods usually shorter than one day, have large RV variability. Among SB1 candidates the majority of objects have low RV variability, and 23 objects have RV variability greater than 140 km/s for different time spans, from one day up to a few years. 

\section{SB1 candidates in the \rave-only sample} \label{Sec:SB1rave}

SB1s were searched for in \rave using the method described above. This is similar to  \citet{matijevic2011} but we applied it to a larger sample. Instead of considering only data obtained up to the third data release we use the sixth and final \rave data release. This increases the number of SB1 candidates from 1333 to 3838 and keeps their percentage at $\sim$10$\%$ of the stars with repeated observations. We note that the fraction of SB1 candidates reaches $\sim 30$~\%\ for objects with a large number of observations (Table \ref{Tab:noobs}). In Table \ref{Tab:listSB1} we report the basic properties of our 3838 SB1 candidates. Their individual RV measurements are published by \citep{steinmetz2019}.

\begin{deluxetable*}{lrrrrr}
\caption{Representative extract from full list of SB1 candidates, reporting the number of observations ($Nobs$), their time-span in days, and the epoch of the first and the last observation by \textsc{Rave}. The whole list of 3838 objects is published electronically.
\label{Tab:listSB1}}      
\tablehead{
Object & $p_{log}$ & $Nobs$ & Time-span & Epoch(first) & Epoch(last) 
}
\startdata
J000107.9-412208 & 4.38 & 2 & 1862 & 2005-08-06 & 2010-09-11 \\
J000349.3-405352 & 4.10 & 3 & 1564 & 2003-08-09 & 2007-11-20 \\ 
J114037.8-260605 & 3.05 & 6 & 123 & 2009-01-27 & 2009-05-30 \\ 
J114110.6-320837 & 6.47 & 7 & 1167 & 2006-03-18 & 2009-05-28 \\ 
J114155.8-235143 & 7.00 & 4 & 100 & 2009-02-19 & 2009-05-30 \\ 
\enddata
\end{deluxetable*}

\begin{figure*}
   \centering
   \includegraphics[width=16cm]{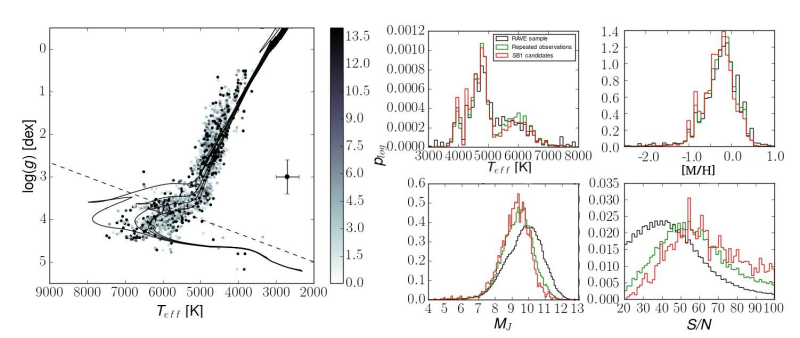}
   \caption{\textit{Left panel}: Kiel diagram of SB1 candidates, color coded according to $ p_{log}$ values. The dashed line separates main sequence dwarfs from giant stars. The Padova isochrones plotted as solid lines have solar metallicity and ages of 1 - 4 Gyr with steps of 1 Gyr. \textit{Right panel}: Histograms of effective temperature (MADERA pipeline), metallicity (MADERA pipeline), magnitude, and signal to noise ratio show distributions that are generally different for the complete \rave sample (black), for objects with multiple observations (green), and for SB1 candidates (red).}
    \label{Fig:kandidati}
\end{figure*}   

\begin{figure}
   \centering
   \includegraphics[width=\columnwidth]{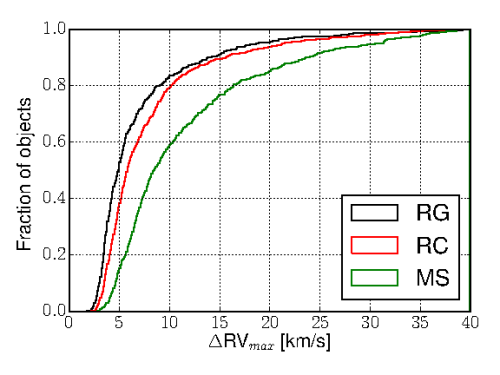}
   \caption{Cumulative histogram of maximum radial velocity changes between measurements for red giant (RG), red clump (RC) and main sequence (MS) SB1 candidates. This diagram shows that RV variability is the largest for MS stars and the lowest for RG stars. MS stars have statistically lower masses, smaller orbits, and consequently larger RV variability in comparison to RG and RC stars.}
      \label{Fig:maksrv}
\end{figure}

Next we study the physical properties of the primary stars in SB1 candidate systems. Figure \ref{Fig:kandidati} shows two peaks in the distribution of the effective temperature, one at $ \sim $ 4500 K for the red clump and giant stars with masses larger than 1.5 \textit{M$ _{\odot} $} and another at $ \sim $ 6000 K for the main-sequence dwarfs with masses $ \sim $1 - 1.2 \textit{M$ _{\odot} $}. SB1 candidates have a slightly lower metallicity than general population, maybe due to a contribution from the secondary star spectrum. The S/N of the re-observed stars and SB1s is higher than in the general population because brighter stars are re-observed more frequently than the faint ones in \rave (so observation time was used more efficiently). The same properties can be seen also in an apparent magnitude histogram. Objects with repeated observations and SB1 candidates have lower magnitudes than the general \rave sample. As shown in Figure \ref{Fig:timespan}, the time-span between the first and the last observation of a given object is around 10 years, so systems with significantly larger orbital periods cannot be detected. This is demonstrated also by Figure \ref{Fig:maksrv}, which shows that the most probable maximum RV differences are $\sim 5$~km~s$^{-1}$, which at our limit of $p_{log} = 2.87$ corresponds to a pair measurements with uncertainties $\sim 1.2$~km~s$^{-1}$, a typical value for \rave. A slight dependence of the position of the most probable RV differences on stellar type is therefore driven by the fact that giants tend to have their RVs measured with a greater precision, as their spectral lines are numerous and sharper. The maximal RV differences can reach 60 or even 100~km~s$^{-1}$, which should correspond to rather close systems with short orbital periods. 

\section{Orbital parameters for main sequence dwarfs} \label{Sec:orbpars}

The radial velocity  of a binary star is given by the following equation:
\begin{equation}
RV = \frac{2\pi a \sin i}{P \sqrt{1 - e^{2}}} \cdot \left[ \cos (\Theta + \omega) + e \cos \omega \right] + \gamma
\end{equation}
where \textit{i} is the orbital inclination, \textit{e} the eccentricity,  \textit{P} the orbital period, $\Theta$ the true anomaly, $\omega $ the longitude of periastron,  $\gamma$ the radial velocity of the center of mass, and  \textit{a} the semi-major axis.  The latter relates to the mass of the primary star (M$_1$) and the mass ratio \textit{q = M$_{2}$/M$_{1}$} as
\begin{equation}
a = \sqrt[3]{\frac{P^{2}GM_{1}(1 + q)}{4\pi^{2}}} \cdot \frac{q}{1+q} 
\end{equation}
All 6 parameters cannot be determined from the small number of re-observations of a given object obtained by \textsc{Rave}, but for objects with at least 4 RV determinations well distributed over time one can attempt a probabilistic approach, with the goal of obtaining approximate  estimates for their orbital periods. To do so one should adopt a grid of parameter values that are to be tested. These are given in (Table \ref{Tab:ranges}). We limit our analysis to primary stars on the main sequence, so that we could infer their mass $M_1$ from their spectroscopically determined effective temperature.  

\begin{deluxetable*}{lcl}
\caption{Ranges of parameters for main sequence dwarfs. \label{Tab:ranges}}      
\tablehead{
Parameter & Range & Step or Values
}
\startdata
Angle of inclination (\textit{i})	& 10 ... 90 deg	& random selection of its sin $ \textit{i} $ value \\
Eccentricity (\textit{e})	& 0.0 ... 0.8	& 0.05 \\
Orbital period (\textit{P})	& 1 ... 3600 days	& 220 logarithmic steps \\
Longitude of the periastron (\textit{$\omega$})	& 0 ... 360	& random selection \\
Mass ratio (\textit{q})	& 0.1 ... 0.85 & 0.05 \\
\enddata
\end{deluxetable*}

\begin{deluxetable*}{lcDDDDDDDD}
\caption{Estimated values of  mass of the primary star (\textit{M$_{1}$}), mass ratio (\textit{q}), orbital period (\textit{P}) in days, eccentricity (\textit{e}), and system velocity (\textit{$\gamma$}) in km~s$^{-1}$.
See text for a discussion of typical errors, in the last four columns the median values, together with lower and upper quartile limits are reported. The whole table is published in electronic form only.}
\label{Tab:parametersSB1} 
\tablewidth{0pt}
\tablehead{
Object & \textit{M}$_{1}/$M$_\odot$ & \multicolumn{2}{c}{\textit{q}}& \multicolumn{2}{c}{\textit{P}}& \multicolumn{2}{c}{\textit{e}}& \multicolumn{2}{c}{\textit{$\gamma$}} }
\startdata
J121104.5-354818	&	1.27	&	0.45&$_{-0.10}^{+0.10}$	&	61&$_{-42}^{+68}$	&	0.25&$_{-0.20}^{+0.30}$ &	-27&$_{-3}^{+2}$		\\
J154304.9-122933	&	1.40	&	0.80&$_{-0.15}^{+0.05}$	&	77&$_{-59}^{+48}$		&	0.20&$_{-0.20}^{+0.30}$ &	8&$_{-3}^{+3}$	\\
J021532.1-363260	&	1.20	&	0.65&$_{-0.15}^{+0.15}$	&	49&$_{-30}^{+15}$	&	0.15&$_{-0.15}^{+0.30}$ &	48&$_{-3}^{+2}$	\\
J142501.1-290222	&	1.30	&	0.80&$_{-0.15}^{+0.05}$	&	4&$_{-2}^{+4}$		&	0.00&$_{-0.00}^{+0.05}$ &	-27&$_{-4}^{+3}$	\\
J093202.1-083428	&	1.05	&	0.20&$_{-0.05}^{+0.15}$	&	535&$_{-262}^{+160}$	& 		0.35&$_{-0.20}^{+0.20}$ & 35&$_{-1}^{+1}$		\\
\enddata
\end{deluxetable*}

\begin{figure}
   \centering
   \includegraphics[width=\columnwidth]{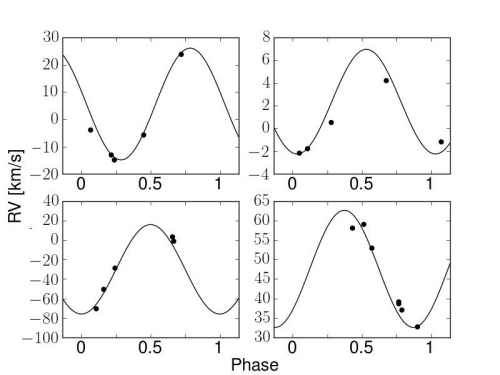}
   \caption{Examples of radial velocity curves for a few binary systems. Black dots are RVs measured by \rave and the curve is calculated with orbital parameters obtained with our method. The objects are J154304.9-122933 (a), J115256.0-161543 (b), J142501.1-290222 (c), and J021532.1-363260 (d).
   }
      \label{Fig:rvfit}
\end{figure} 

\begin{figure*}
   \centering
   \includegraphics[width=16cm]{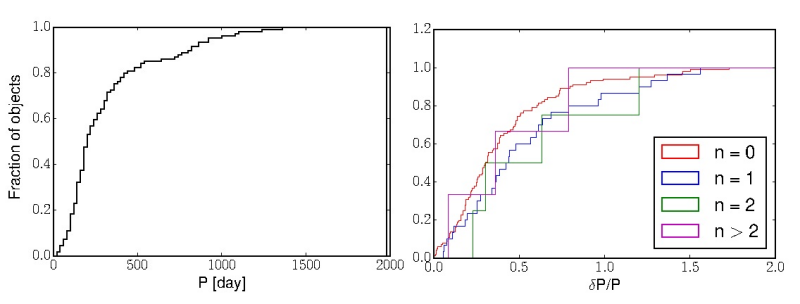}
   \caption{\textit{Left panel:} Cumulative diagram of estimated orbital periods. Most of the SB1 systems have periods shorter than 2 years. \textit{Right panel:} Cumulative period dispersion distributions for different trends in radial velocity changes.
   }
      \label{Fig:Apmedianraspr}
\end{figure*}

\begin{figure*}
   \centering
   \includegraphics[width=16cm]{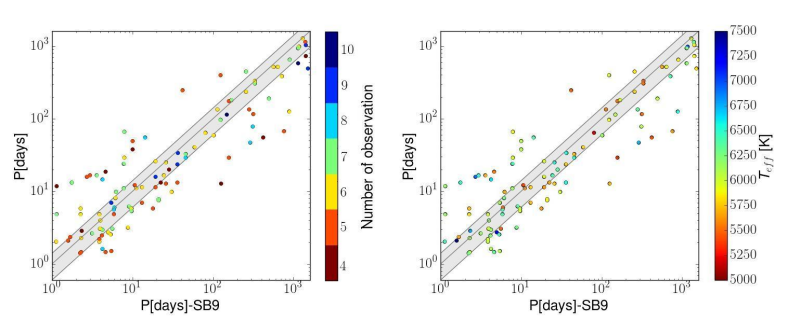}
   \caption{Comparison of true period values for objects in the 9th Catalogue of Spectroscopic Binaries and for periods calculated with the method described in Section \ref{Sec:orbpars}. The grey area represents offsets from unity of $\pm$ 40 $\%$ of the period value. The left panel is color coded by the number of observations and the right panel by the primary star's temperature. It seems that neither temperature nor the number of observations have a significant impact on our results.}
\label{Fig:pourbaix}
\end{figure*}

All combinations of these parameters do not occur in nature. Following \citet{duquennoy1991} and \citet{raghavan2010} we adopt the following constraints: 
\begin{enumerate}
\item main-sequence systems with $P<12$~days are assumed to have circular orbits due to tidal interaction in close binaries;
\item systems with a period of $P >12$~days have a flat distribution in eccentricity that varies from 0.0 to $ \sim  0.8$, independent of period; and
\item short-period systems ($P< 100$~days) have $q > 0.4$.   
\end{enumerate}

There are 406 SB1 candidates with primaries that are main-sequence dwarfs and that have at least 4 RVs measured by \rave. This makes them suitable objects to attempt an approximate determination of their orbital periods. To do so, we first split them into 4 groups according to the sign of their RV changes: in the first group are objects where the RV derivative is either positive or negative throughout (n = 0), and in the other three groups are objects with one, two, or more changes in the sign of their RV derivative (n = 1, 2, $> 2$). The first two groups are well populated while there are only a handful of objects in the last two groups. These groups can be used to roughly infer what are likely values of $P$. For the first two groups (n = 0, 1) we assumed that $P$ cannot be longer than 8 times the time-span between the first and the last observation, for the third group (n = 2) we lower the maximum period to 3 times the time-span and for the last group (n~$ >2$) to within the time-span. Next we use these limits on $P$ to compare observed RVs to the calculated ones by marginalizing over other parameters. In particular, for each object we generate 500 sets of randomized values of inclination, longitude of periastron, and initial orbital phase and for each of these sets we check on all allowed combinations of period, eccentricity, and mass ratio, as quoted in Table 3. 

The number of parameters is similiar or even larger than the number of data points available, so we are not in a position to determine their values. Still, likely ranges of some parameters, such as mass ratio, orbital period, system velocity and a rough estimate of eccentricity can be judged from our data, so we report these estimates in Table 4. On the other hand the initial orbital phase, longitude of periastron, and inclination are either completely driven by uncertainties in orbital period or they are poorly constrained by the RV nature of our data -- so we refrain from reporting them in Table 4. Typical errors on mass of the primary star are determined to within  10\% using the effective temperature of the primary and its assumed position on the main sequence. For other values their median value and quartile brackets are reported. The uncertainties are substantial, still we believe such information is useful. For example, it is clear that J142501.1-290222 is a binary with an orbital period of just a few days and with quite similar masses of the components. The system ($\gamma$) velocity is relatively well constrained from derived sets of orbital solutions, so these values may be useful when estimating  Galactic orbits of SB1 candidates. Figure \ref{Fig:rvfit} illustrates solutions for a few objects. 

An SB1 with a very high eccentricity can be very hard to identify. Such a binary spends only a very short time close to periastron at high orbital velocities, but most of the time their RVs are nearly constant and close to the $ \gamma $ velocity of the center of mass. \rave objects generally have a small number of observations (Figure \ref{Fig:noobs}), so it is quite likely to miss the RV spike around the periastron passage and the object might not be identified as a binary at all.

Most of the SB1 candidates have orbital periods shorter than a year, and only $\sim$ 10$\%$ of the objects have orbital periods larger than 2 years (Figure \ref{Fig:Apmedianraspr}, left panel). As already mentioned, orbital periods cannot be very long as observations span only a few years or less, but they cannot be very short either as large RV variations are quite rare. The right panel of Figure \ref{Fig:Apmedianraspr} shows that the orbital period is determined to within 50\%\ for about half of the objects. Similar uncertainties are true also for $q$.

In order to better evaluate the precision and reliability of our orbital period determinations we constructed a mock sample of RV measurements using orbital solutions of SB1 binaries as reported in the $9^{th}$ Catalogue of Spectroscopic Binaries \citep{pourbaix2004}. For each \rave SB1 candidate we selected its counterpart in the catalog, matching the effective temperature and dwarf nature of its primary. The orbital phase of the first \rave observation was picked randomly, with subsequent velocities generated from the orbital solution at the same time-offsets as in the actual \rave observations. The set was processed in the same way as the \rave observations, so that the derived orbital period could be compared to the actual one from the catalog (Figure \ref{Fig:pourbaix}). 50 $\%$ of all objects are in the gray area, where the dispersion is $\pm$ 40 $\%$ around the true value, while 90 $\%$ of all objects have periods determined to within a factor of 2. The results do not depend on the effective temperature of the primary (see right panel in Figure \ref{Fig:pourbaix}). The results are generally acceptable but we note some systematics: our periods tend to be overestimated for short period binaries and underestimated for long-period ones.

\section{SB2 candidates} \label{Sec:sb2s}

\begin{figure}
   \centering
   \includegraphics[width=\columnwidth]{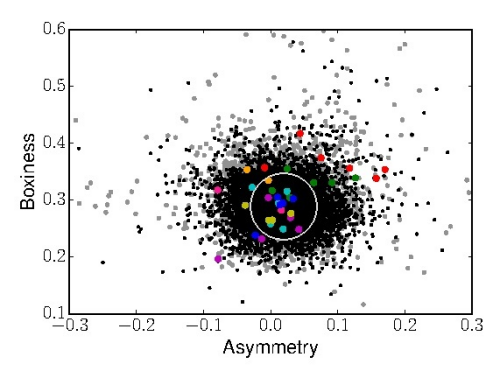}
   \caption{Asymmetry and boxiness of the calcium triplet for single stars (black dots), SB1 candidates (gray dots), and 8 SB2 candidates, each in a different color. SB1 candidates with at least one spectrum outside the marked circle with a radius of 0.05 centered on (0.01, 0.3)  were visually inspected for the presence of secondary light in their spectra. 
   }
   \label{Fig:boxasy}
\end{figure} 

\begin{figure}
   \centering
   \includegraphics[width=\columnwidth]{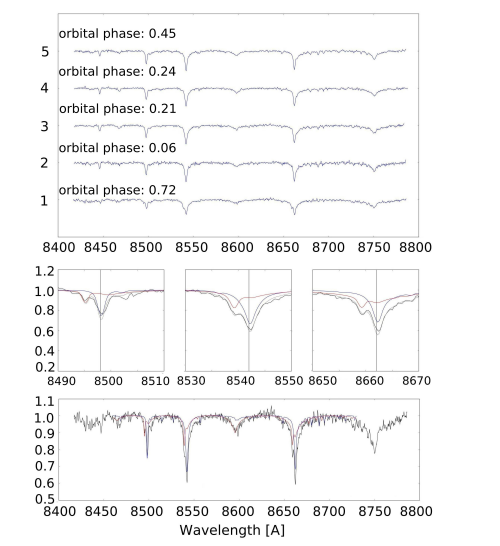}
   \caption{Top panel: spectra of the SB2 candidate J154304.9-122933 with its orbital phases labelled. The RV curve of this object is shown in Figure \ref{Fig:rvfit}a. Middle panel: Fitted calcium triplet lines for the orbital phase 0.72. Bottom panel: Fitted spectra for the same phase, using 
$RV_1 = -95$~km/s, $RV_2 = 10$~km/s, $T_1 = 7500$~K, $T_2 = 6750$~K, $q = 0.55$, [Fe$/$H]$ = 0.5$.
   }
   \label{Fig:sb2fit}
\end{figure} 

\begin{figure*}
   \centering
   \includegraphics[width=16cm]{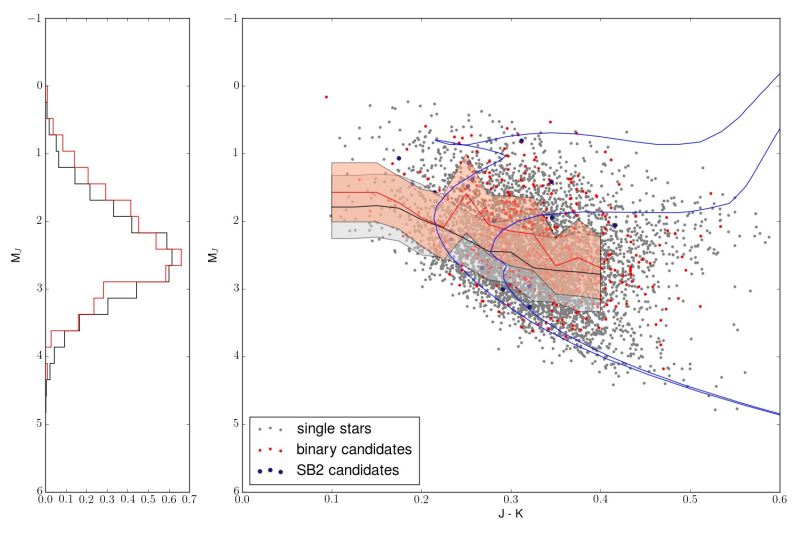}
   \caption{Absolute magnitude - color diagram for single, SB1, and SB2 candidate stars. The blue lines are 2 and 4 Gyr Padova isochrones \citep{marigo2017} with a metallicity set to the Solar value. Gray dots are \rave single stars, red ones are SB1 candidates, and black ones SB2 candidates. The black solid line is the median value of the J magnitude for single stars and the gray area indicates the $ \sigma $ around median. Similarly, red solid line and shaded area are medians for SB1 candidates. 
   }
   \label{Fig:hrsb2}
\end{figure*} 

\begin{deluxetable*}{lccccccc}
\caption{ Most probable values of parameters for double lined binary candidates: mass ratio (\textit{q}), temperature of primary (\textit{$T_1$}), temperature of secondary star (\textit{$T_2$}), and metallicity [Fe$/$H].} 
\label{Tab:SB2}       
\tablehead{
Object & \textit{q} & \textit{$T_1$} & \textit{$T_2$} & [Fe$/$H] 
}
\startdata
J120432.1-203723	&	0.5	&	7250	&	6750 & 0.5\\
J203415.1-201303	&	0.6	&	6750	&	6500 & -0.5\\
J090701.4-142256	&	0.65&	4750	&	4500 & 0.0\\
J125113.4-202156	&	0.5	&	7500	&	6750 & -0.5\\
J154304.9-122933	&	0.55 &	7500	& 	6750 & -0.5\\
J161301.0-130342	&	0.65 &	5000    &	4750 & -0.5	\\
J100235.9-093818	&	0.55 &	7000    &	6250 & -0.5	\\
J045419.4-030709	&	0.55 &  6750	&	6000 & -0.5	\\
\enddata
\end{deluxetable*} 

\quad 
SB1 candidates in \rave have been labelled as normal single stars by a morphological classification scheme \citep{matijevic2012}. This is understandable, as the primary star is usually much brighter than the secondary one in such systems. But a close inspection of their spectra sometimes reveals a contribution of the light from the secondary, thus moving the object to the SB2 category. The signature of the secondary can be searched for in spectral lines. Due to the limited S/N ratios of the \rave spectra the strongest lines turn out to be the most appropriate ones. So we focused on the calcium triplet lines and measured their boxiness and asymmetry indices (Figure \ref{Fig:boxasy}). Boxiness is defined as the ratio of line widths at three quarter and one quarter of the line depth, while asymmetry is the shift of the centroid at half maximum in units of full width at half maximum (FWHM). The reported values are averages over the three calcium lines. For a symmetric Gaussian line the boxiness equals 0.456 and the asymmetry is zero. Calcium triplet lines have broad wings, so one expects somewhat smaller values of boxiness, but the asymmetry should still be very close to zero for a spectrum of a single star, with an exception of hot stars where the presence of Paschen lines of hydrogen contaminates wings of the calcium lines.

 The combination of these values points to spectra with unusual shapes of spectral lines that could be double lined binaries (colored dots in Figure \ref{Fig:boxasy}). Visual inspection of those spectra confirmed the existence of several double lined binaries where a chi-square fit with two spectra gave a much better representation of the observed spectrum than a single spectrum. List of double-lined binaries is given in the \ref{Tab:SB2}.
 
 An example is given in Figure \ref{Fig:sb2fit}. The top panel shows five spectra of the same star, where only the last  one, observed at orbital phase 0.72, shows obvious double components in the calcium  triplet and also in other spectral lines. The RV curve of the same object is presented in Figure \ref{Fig:rvfit}a. The first four spectra in Figure \ref{Fig:sb2fit} were obtained in 2009, while the last one which shows double-lined spectral lines, was obtained in 2006. The bottom two panels show results of a least-square fit to this double-lined spectrum using both RVs, temperatures, mass ratio, and metallicity as free parameters, with surface gravity constrained by the assumption that both stars are on the main sequence. We note that the same solution presents a good fit also for the other four spectra, though one would expect more pronounced double-lined profiles also at the other quarter phase. This may be explained by the fact that we adopted orbital periods and phases as calculated in the SB1 fit, even though a contribution from a secondary component in SB2s may alter these values. The goal of this analysis is to point to possible SB2 candidates, but the number of multiple \rave  spectra is too small to attempt a complete solution anyway. We also note that our list of SB2 candidates contains only 8 objects -- these are the ones that escaped detection by the automated morphological classification algorithm \citep{matijevic2012}. On the other hand, \citet{steinmetz2019} lists 2861 objects with (some) spectra in the SB2 category, with physical properties of 123 of them discussed already by \citet{matijevic2010}.
 
For an object on the main sequence one could expect a moderate increase of luminosity if the object is not single but a SB1 binary system, with an effect even more pronounced for SB2s. This is an obvious consequence of an increasing contribution of light from the secondary. Indeed, the SB1s in Figure \ref{Fig:hrsb2} are about 0.2 ~mag brighter than single stars, but the dispersion around the median value is the same. The data pool of SB2s is too small to make the same statistics, but it can be seen that they are brighter than single stars and some of them approach a 0.75~mag limit, which corresponds to the joint luminosity of two equal stars instead of one. The positions of evolutionary tracks of Solar-type stars in Figure \ref{Fig:hrsb2} demonstrate, however, that many of the single or binary objects may be actually evolving off the main sequence, which also makes these objects brighter. This matter is discussed in \citet{cotar2019}.

\section{Combining \rave with \textsc{Gaia}} \label{Sec:raveandgaia}

\begin{figure}
   \centering
   \includegraphics[width=\columnwidth]{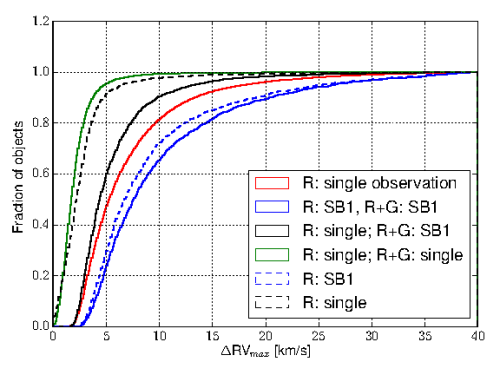}
   \caption{Maximum RV variability for groups of objects identified as single stars or SB1 candidates by \rave DR6 survey alone (label R), or by a combination of the \rave and  \textsc{Gaia} DR2 (label R$+$G) surveys. The red line denotes RV differences between \rave and \textsc{Gaia} for all objects that have only one \rave observation.  
   The green line shows that the RVs of single stars match to $\sim 2$~km~s$^{-1}$ across the surveys, so one can use a combination of RVs from both surveys to search for SB1 candidates. The black solid line demonstrates that a combination of the two surveys can use its long time-span to identify SB1s with the lowest RV amplitude.
   }
   \label{Fig:maksrvgrstep}
\end{figure}

\begin{figure}
   \centering
   \includegraphics[width=\columnwidth]{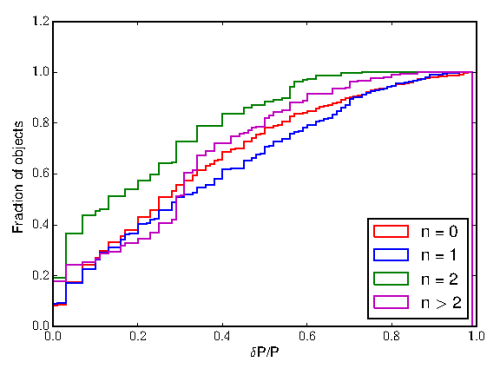}
   \caption{Cumulative histogram of the relative dispersion of period median values for SB1 candidates with \rave DR5 $+$ \textsc{Gaia} DR2 velocities combined. The results are similar to the ones with \rave DR6 data only (Figure \ref{Fig:Apmedianraspr}). About 70\% of the objects have $P$ determined to within 50\%.}
   \label{Fig:medianpgaia}
\end{figure}

\begin{deluxetable*}{lrrrrr}
\caption{Representative sample from full on-line list of SB1 candidates obtained by a combination of \rave and \textsc{Gaia} RV measurements. The table is similar to Table  \ref{Tab:listSB1} but it contains a much larger list of 27,716 SB1 candidates. The complete table will be published only  electronically. An epoch of 15 June 2015 for the \textsc{Gaia} observations was adopted (see Section 7 for an explanation).
\label{Tab:listwithGaiaSB1}}      
\tablehead{
Object & $p_{log}$ & $Nobs$ & Time-span & Epoch(first) & Epoch(last)   
}
\startdata
J012955.0-623622    &   5.99    &   3   &   2462   &   2008-09-17   &    2015-06-15 \\
J135416.6-222607    &   4.30    &   3   &   3758   &   2005-03-01   &    2015-06-15 \\
J004904.6-222139    &   8.91    &   3   &   4262  &   2003-10-14    &    2015-06-15 \\
J070727.9-480148    &   11.56    &   4   &  3126  &   2006-11-23    &    2015-06-15 \\
J051516.8-324737    &   5.53    &   3   &   3423   &   2006-01-30   &   2015-06-15 \\
\enddata
\end{deluxetable*}

\begin{deluxetable*}{lDDDDDDDD}
\caption{Estimated values of mass ratio (\textit{q}), orbital period (\textit{P}) in days, eccentricity (\textit{e}), and system velocity (\textit{$\gamma$}) in km~s$^{-1}$ with lower and upper quartile limits for the combination of the RAVE and the Gaia surveys . The whole table is published in electronic form only.}
\label{Tab:parametersSB1Gaia}
\tablewidth{0pt}
\tablehead{
Object & \multicolumn{2}{c}{\textit{q}}& \multicolumn{2}{c}{\textit{P}}& \multicolumn{2}{c}{\textit{e}}& \multicolumn{2}{c}{\textit{$\gamma$}} }
\startdata
J121104.5-354818	&	0.45&$_{-0.10}^{+0.15}$	&	92&$_{-73}^{+52}$	&	0.25&$_{-0.25}^{+0.35}$ &	-19&$_{-2}^{+2}$		\\
J154304.9-122933	&	0.80&$_{-0.15}^{+0.05}$	&	77&$_{-59}^{+48}$		&	0.40&$_{-0.35}^{+0.25}$ &	6&$_{-4}^{+3}$	\\
J021532.1-363260	&	0.65&$_{-0.15}^{+0.15}$	&	49&$_{-30}^{+15}$	&	0.15&$_{-0.15}^{+0.25}$ &	49&$_{-2}^{+3}$	\\
J142501.1-290222	&	0.80&$_{-0.15}^{+0.05}$	&	4&$_{-2}^{+4}$		&	0.00&$_{-0.00}^{+0.05}$ &	-33&$_{-4}^{+4}$	\\
J093202.1-083428	&	0.20&$_{-0.05}^{+0.20}$	&	341&$_{-107}^{+257}$	& 		0.45&$_{-0.25}^{+0.15}$ & 35&$_{-1}^{+1}$		\\
\enddata
\end{deluxetable*}

The \textsc{Gaia} satellite was launched on 2013 December 19 and started with scientific observations in 2014 July \citep{prusti2016}. Its main objectives are astrometric measurements of parallax and proper motion, but here we focus on results from an onboard RV spectrometer. It has a resolving power of $\sim$~11,500 covering the near-infrared wavelength range at 845 - 782 nm, with an expected precision of 1~km~s$^{-1}$  for GK stars brighter than $G \sim 12$ \citep{cropper2018,katz2019}. In 2018 April the \citet{brown2018} published median RVs for 7.2 million sources with effective temperatures in the range of 3550--6900~K that are brighter than $G = 12.5$. These medians were obtained over 22 months of observations (July 25 2014 - May 23 2016), they have a typical overall precision of 1.05~km~s$^{-1}$. Most of the stars observed also by \rave lie at the faint end of objects accessible for \textsc{Gaia} RV measurements. For such objects ($G \sim 11.8$) the precision of \textsc{Gaia} RVs is 1.4 and 3.7~km~s$^{-1}$ for an effective temperature of 5000~K and 6500~K, respectively. We note that these errors are not significantly larger than for the \rave survey, so the two sets of measurements can be efficiently combined.  Indeed, for a vast majority of objects labelled as single stars in RAVE their \textsc{Gaia} RV is matching closely (green solid line in Figure 11). This demonstrates that a small zero point offset between these two datasets  \citep[estimated at 0.3~km~s$^{-1}$, ][]{katz2019} does not influence our analysis when our relatively stringent limits on $p_{log}$ used to search for SB1 candidates are considered. \textsc{Gaia} DR2 published median value of RV and its dispersion only for objects judged to have a constant RV during the 22 months of observations. Any objects with a pronounced RV variation or SB2s were  excluded. So we can expect to identify only objects with low amplitudes of RV variation which are typical for objects with long orbital periods (black solid line in Figure 11 demonstrates that this is indeed the case). Since their orbital periods are generally much longer than the 22-month span of \textsc{Gaia} observations it is safe to assume that all \textsc{Gaia} measurements were obtained at a similar orbital phase. If this was not the case one would have to consider effects of averaging velocities at different orbital phases which pushes their median close to system velocity. This effect is ignored in our analysis. Note that by doing so we may make a moderate underestimate of detected new SB1s, as a measurement close to system velocity dumps detected amplitudes of RV variation.

\textsc{Gaia} obtained its RV measurements after completion of the \rave survey, so combining the two datasets increases the time-span of RV measurements and thus allows a detection of SB1 systems with longer orbital periods. More importantly, most of its targets have been observed with \rave only once (Figure \ref{Fig:noobs}), but \textsc{Gaia} is adding another observation and so allows to test for variability of their RVs. In fact,  \textsc{Gaia} itself observed each object several times, but these observations will be published only in one of the next data releases and they do not reach the combined timespan of \rave $+$ \textsc{Gaia}, which stretches up to over a decade (Figure \ref{Fig:timespan}). 

Median \textsc{Gaia} RVs have been calculated from several observations over 22 months. Individual measurements are not available, so we adopted an epoch of 15 June 2015, which is at the middle of the observed timespan. Note that the true median epoch of \textsc{Gaia} RV measurements could be up to a few months earlier or later. But this has little influence on our analysis, as the closest \rave observations were obtained at least 2.1 years earlier. The combination of \rave and \textsc{Gaia} RVs extends the maximum time span from 8 to 12 years, while the median value is extended from 2 to 7 years (Figure \ref{Fig:timespan}).
We note that the use of median velocity of \textsc{Gaia} favours an analysis which is separate from the one based on \textsc{Rave} data only

There are 450,646 stars with RVs measured in both surveys. This is a remarkable increase by over 37,661 stars suitable for SB1 search in the \textsc{Rave}-only survey. After conducting the same analysis as before (Section \ref{Sec:method}) we obtained the following results. 7.7\%\ of the stars with observations in both datasets are SB1 candidates. This is close to a the fraction of 10.2\%\ that was obtained based on \rave data only, even though the time-span is much longer. Among \rave stars with multiple observations that were labeled as single stars ($ p_{log} $ $<$ 2.87) we found almost 10 \%\  new binary candidates after we included \textsc{Gaia} velocities in calculations. Overall, we were able to identify 27,716 SB1 candidates in the \rave $+$ \textsc{Gaia} sample, compared to 3838 from the \textsc{Rave}-only analysis.

The black dashed line  in Figure \ref{Fig:maksrvgrstep} shows the radial velocity variability for stars that were classified as normal single stars, according to their \rave radial velocities. For the vast majority of objects the RV changes are less than 5 km~s$^{-1}$, so we can assume those objects are long period binaries, impossible to detect without observations over a longer time-span. After we added \textsc{Gaia} velocities, the resulting RV variability became significant for some of these stars and we identified new SB1 candidates. Future \textsc{Gaia} data releases will probably reveal even more binary candidates.    

Next we repeated the computation of orbital parameters (Section \ref{Sec:orbpars}), now adding \textsc{Gaia} velocities. The results are shown in Figure \ref{Fig:medianpgaia}. The results are very similar as for the \textsc{Rave}-only dataset. This is a consequence of the similar accuracy of RV measurements in both samples. Median values together with lower and upper quartile limits are reported in the \ref{Tab:parametersSB1Gaia}.  More than 90\%\ of objects have all of the derived parameters with combination of the Rave and \textsc{Gaia}  velocities in the interquartile range (between upper and lower quartiles) of \textsc{Rave}-only dataset.

\section{Conclusions} \label{Sec:conclusions}

This paper presents a complete list of SB1 candidate stars in the \rave survey based on the requirement that their RV measurements differ by at least $\sim 4.2 \sigma$ apart ($p_{log} = 2.87$).  Using the probability function described in \citep{matijevic2011} we detected 3838 single lined spectroscopic binary candidates. This almost triples the number of candidates known so far and corresponds to  $\sim 10 $ \%\ of all normal stars observed multiple times by \textsc{Rave}. Most of the primary stars of these systems belong to main-sequence dwarfs with temperatures around 6000 K and masses around 1 - 1.2 M$_{\odot}$, or red clump stars and red giant stars with temperatures 4500 - 5000 K and masses larger than 1.5 M$_{\odot}$. The secondary stars contribute only a small fraction of the total light of an SB1 candidate. Still, the spectral lines in the combined spectrum are somewhat shallower, so this may be the reason why SB1 candidates appear to be more metal poor than the general \rave population.   

Even though most of the stars with repeated observations in \rave have been observed only a few times, it is possible to make a rough estimate of the orbital parameters for systems with primary components on the main sequence. We focused on systems with at least 4 observations. Being limited by the time span between re-observations, our results showed that most systems have an orbital period shorter than one year, and only a few of them have orbital periods of around three years. 

Our sample of SB1 candidates includes stars morphologically classified as normal single stars. But at least in some cases one may hope to identify a contribution of the secondary component to the total light of the system. In the spectra this is revealed by unusual shapes of the spectral lines, which were measured through the boxiness and asymmetry of calcium triplet lines. A visual inspection of their spectra revealed some compelling cases with a mass ratio around 0.8. We also note that both SB1 and SB2 candidates tend to be somewhat brighter than their single-star counterparts, which is consistent with a contribution of light from a secondary component.

\textsc{Gaia} DR2 is supplementing the \rave dataset with another RV observation for 450,646 stars. It also observed at an epoch after \rave observations were concluded, so the combined dataset has a larger time-span of up to 12 years and with a median of 7 years. The analysis of the combined datasets allows us to identify 27,716 stars as single lined binary candidates, which presents an order of magnitude increase over earlier studies. The orbital and physical properties of these systems are similar to the ones from the \textsc{Rave}-only dataset, but an accurate knowledge of their spatial position and velocity vectors provided by \textsc{Gaia} DR2 allows us to calculate their Galactic orbits and to further characterize their physical parameters. 

The contents of Tables \ref{Tab:listSB1}, \ref{Tab:parametersSB1}, \ref{Tab:listwithGaiaSB1}, and \ref{Tab:parametersSB1Gaia} are available in electronic edition.  

\acknowledgments
We thank the anonymous referee for useful comments which improved the clarity of the manuscript. Funding for \rave\  has been provided by: the Leibniz-Institut f\"{u}r Astrophysik 
Potsdam (AIP); the Australian Astronomical Observatory;  the Australian National 
University; the Australian Research Council; the French National Research Agency; the 
German Research Foundation (SPP 1177 and SFB 881); the European Research Council 
(ERC-StG 240271 Galactica); the Istituto Nazionale di Astrofisica at Padova; The Johns 
Hopkins University; the National Science Foundation of the USA (AST-0908326); the W. M. 
Keck foundation; the Macquarie University; the Netherlands Research School for Astronomy; 
the Natural Sciences and Engineering Research Council of Canada; the Slovenian Research 
Agency (core funding No.\ P1-0188); the Swiss National Science Foundation; the Science \& Technology Facilities 
Council of the UK; Opticon; Strasbourg Observatory; and the Universities of Basel, Groningen, 
Heidelberg and Sydney. TZ thanks the Research School of Astronomy \& Astrophysics in Canberra for support through a Distinguished Visitor Fellowship.

This work has made use of data from the European Space Agency (ESA) mission
{\it Gaia} (\url{https://www.cosmos.esa.int/gaia}), processed by the {\it Gaia}
Data Processing and Analysis Consortium (DPAC,
\url{https://www.cosmos.esa.int/web/gaia/dpac/consortium}). Funding for the DPAC
has been provided by national institutions, in particular the institutions
participating in the {\it Gaia} Multilateral Agreement.

\end{document}